\documentclass{Rinton-P9x6}
\usepackage{amsmath}
\usepackage{graphicx}
\usepackage{amsfonts}

\begin{document}

\newcommand{\ket}[1]{\left| #1 \right>}
\newcommand{\bra}[1]{\left< #1 \right|}

\title{Shortening Grover's search algorithm for an expectation value quantum computer}
\author{David Collins}

\address{Department of Physics, Carnegie Mellon University, 
             Pittsburgh PA 15213\\
         Email: collins5@andrew.cmu.edu}

\maketitle

\abstracts{Quantum algorithms are conventionally formulated for implementation on a single system of qubits amenable to projective measurements. However, in expectation value quantum computation, such as nuclear magnetic resonance realizations, the computer consists of an ensemble of identical qubit-systems amenable only to expectation value measurements. The prevalent strategy in such expectation value implementations of quantum algorithms has been to retain the conventional formulation's unitary operations but modify its initialization and measurement steps appropriately. This naive approach is not optimal for Grover's algorithm and a shortened version for expectation value quantum computers is presented.}

Quantum algorithms~\cite{NC} are usually formulated with the idea that each will be implemented on single quantum system amenable to projective measurements. In contrast, most experimental implementations have used room temperature solution state NMR, which involves an ensemble of non-interacting identical quantum computers and for which the only measurement outcomes available are expectation values of observables~\cite{ChGKL,VSSYBCh00,vandersypen01}. Consequently quantum algorithms have been adapted for such expectation value quantum computers by modifying the preparation and measurement stages~\cite{ChGKL,cory98}. However, these adaptations have been constructed so that the expectation value quantum computer mimics its single quantum system relative as closely as possible. As such, they do not use the multiple identical quantum computers in the ensemble advantageously. In this article we show how the ensemble can be used to shorten Grover's search algorithm for expectation value quantum computers (a more detailed description appears elsewhere~\cite{collins02}). 

We consider an expectation value quantum computer, consisting of an ensemble of identical, non-interacting quantum computers, which (i) can be prepared in any pure or pseudopure initial state\cite{ChGKL,cory98}, (ii) to which any unitary transformation can be applied, and (iii) for which the expectation value (EV) of $\sigma_z(k)$ can be measured for each qubit (labeled by $k$). This is the case with current NMR realizations.

The standard translation of a quantum algorithm to a form suitable for an expectation value quantum computer can be illustrated via Grover's algorithm for searching a database containing a single marked item in one of $N$ possible locations (for convenience assume that $N=2^L$ for some integer $L$; this is not a severe restriction). The database will be represented as $X=\{ 0,1, \ldots , N-1 \}$ and the search proceeds with the aid of the following oracle, invoked at unit cost: $f(x)=0$ if $x \neq s$ and $f(x)=1$ if $x= s$ where $s$ denotes the marked item's location.  A classical sequential search requires $N/2$ oracle queries on average to locate the marked item correctly. A quantum algorithm~\cite{NC,grov} uses an $L$ qubit data register and an oracle unitary operation, defined on the computational basis states as $\hat{U}_f \ket{x} := (-1)^{f(x)} \ket{x}$ and extended linearly to superpositions of these. The data register is initialized to $\ket{\psi_i} :=1/\sqrt{2^L}\sum_{x=0}^{2^L-1} \ket{x}$, followed by repeated applications of the Grover iterate $\hat{G} := \hat{D} \hat{U}_f.$ Here $\hat{D} \left( \sum_{x=0}^{N-1} c_x \ket{x} \right) := \sum_{x=0}^{N-1} \left( -c_x + 2 \left< c \right> \right) \ket{x},$ where $\left< c \right> =\sum_{x=0}^{N-1} c_x / N$. After $m$ applications of $\hat{G}$, the data register's state is~\cite{boyer98} 
\begin{equation}
\ket{\psi} =    \sin{\left[ (2m+1)\theta/2\right]} \ket{s} + 
                \frac{\cos{\left[ (2m+1)\theta/2\right] }}{\sqrt{N-1}} \sum_{x \neq s} \ket{x}
\label{eq:gen_grover_state}
\end{equation}
where $\cos{\theta} = 1-2/N.$ After $\lfloor\pi/2\theta\rfloor$ applications of $\hat{G}$  the system's state is approximately $\ket{\psi} = \ket{s}$;  a projective measurement (PM) in the computational basis would yield $s$ with probability of error, which we henceforth ignore, at most $1/N$. For $N \gg 1,$ $\theta \approx 2\sqrt{1/N}$ and the algorithm requires about $\lfloor \sqrt{N}\; \pi/4 \rfloor$ oracle invocations to locate the marked item with near certainty, giving a quadratic speedup over the classical sequential search. To convert between PM and EV outcomes note that, when $\ket{\psi} = \ket{s}$, $\left< \sigma_z(k) \right> := \bra{\psi} \sigma_z(k) \ket{\psi} = (-1)^{s_k}$ where $s= s_L \ldots s_1$ is the binary representation of $s$. This reveals the marked item if it is possible to distinguish between $\left< \sigma_z(k) \right> = +1$ and $\left< \sigma_z(k) \right> = -1$ for each qubit. This \emph{standard EV version} of Grover's algorithm has been used in all NMR realizations to date.

What is crucial for the success of the above translation, is that the sign of $\left< \sigma_z(k) \right>$ for each qubit can be determined. This depends only on which PM outcome would be more likely: positive or negative for PM outcomes $0$ or $1$ respectively. Therefore, for expectation value quantum computers, it is not essential that the state prior to measurement be $\ket{s}$, but rather one for which the most likely PM outcome is $s$. This can be attained with $m <\lfloor \sqrt{N}\; \pi/4 \rfloor$ applications of $\hat{G},$ in which case, Eq.~(\ref{eq:gen_grover_state}) yields
\begin{equation}
  \left< \sigma_z(k) \right>  =  (-1)^{s_k} A_m
 \label{eq:trunc_grover_ev}
\end{equation}
where the \emph{EV attenuation}
\begin{equation}
  A_m := \frac{\sin^2{[(2m+1)\theta/2]}\; N - 1}{N-1}
 \label{eq:amp_after_m}
\end{equation}
determines the magnitude of the EV. The EV attenuation increases monotonically with with respect to $m$ from $0$ prior to any applications of $\hat{G}$ to approximately $1$ in the case of the standard EV version.  This leads to a \emph{truncated EV version} of Grover's algorithm in which is identical to the standard EV version \emph{except that it terminates after the minimum number of applications of $\hat{G}$ such that it is still possible to distinguish reliably between $\left< \sigma_z(k) \right> >0$ and $\left< \sigma_z(k) \right> < 0$ for all data register qubits.} In an ideal case (i.e.\ an infinitely large ensemble and infinitely precise measurements) the truncated EV version succeeds after one application of $\hat{G}.$ In practice, noise and statistical limitations effectively establish a threshold EV attenuation, $A_{\textrm{th}}$, below which the sign of the EV cannot be discerned reliably; the algorithm then succeeds when $A_m > A_{\textrm{th}}$. The threshold will depend on experimental details but, for many practical expectation value quantum computers it will possible to truncate the algorithm compared to the standard EV version. 

Searching a database containing more than one marked item is complicated by the fact that the standard version of Grover's algorithm is not deterministic. Here, after about $\lfloor \sqrt{N/M}\; \pi/4 \rfloor$ applications of $\hat{G}$, the data register is approximately in the state $\ket{\psi_f} = 1/\sqrt{M} \sum_{x \in S} \ket{x}$ where M is the number of marked items and $S$ denotes the set their locations~\cite{NC}. A PM will yield a location of one of the marked items with certainty. However, the bitwise EV technique described earlier may fail in some cases since averaging over more than one possible database location can produce devastating cancellations~\cite{collins02}. One way to avoid this and still incorporate the truncation described above is to use a \emph{filtered EV} technique based on iterated runs of the algorithm. The iteration step produces the $k+1$ bit of a marked item's location if the first $k$ bits have been determined to be $s_k, s_{k-1}, \ldots , s_1$ by combining the EVs of two runs of the algorithm. In the first the algorithm is run without modification, while in the second a \emph{correlation operation} $\hat{C}_{k+1}(s_k,\ldots,s_1)$ is inserted between the last application of $\hat{G}$ and the EV measurements, where
\begin{equation}
 \hat{C}_{k+1}(s_k,\ldots,s_1) \ket{x_L \ldots x_1} :=
 \left(\sigma_x(k+1) \right)^{g(x_k\ldots x_1)} \ket{x_L \ldots
 x_1},
\end{equation}
with
\begin{equation}
 g(x_k\ldots x_1) := (x_k \oplus s_k \oplus 1)(x_{k-1} \oplus s_{k-1} \oplus 1)\ldots  (x_1 \oplus s_2 \oplus 1) \oplus 1.
 \label{eq:corr_function}
\end{equation}
Averaging the EVs obtained from the two runs gives~\cite{collins02}
\begin{equation}
 \left< \sigma_z(k+1) \right>_{\textrm{ave}} = \frac{1}{M} 
                                   \sum_{x \in S'}
                                   (-1)^{x_{k+1}},
\end{equation}
identical to an EV performed on $1/\sqrt{M} \sum_{x \in S'} \ket{x}$ where $S':=\{ x \in S: x_k = s_k,  \ldots ,x_1=s_1\}$, and whose sign gives $s_{k+1}$ for some marked item whose  first $k$ bits are $s_k, s_{k-1}, \ldots ,s_1$. The process is initiated by inspecting the sign of $\left< \sigma_z(1) \right>,$ giving $s_1$ for a marked item. The entire filtering procedure requires $\log_2{N}$ runs of the algorithm and $O((\log_2{N})^2)$ operations to compute $g$ for all stages. Note that the amplitude of the EV scales as $1/M$ and thus for a given experimental realization there will be a threshold (in $M$) beyond which this scheme fails. We only consider cases where $M$ lies beneath this threshold.

The truncated EV version of Grover's algorithm for one marked item is extended to multiple marked items if the filtering scheme described above is used. After $m$ applications of $\hat{G},$ 
\begin{equation}
 \left< \sigma_z(k) \right>_{\textrm{ave}}  =  \frac{A_m}{M} \sum_{x \in S'} (-1)^{x_k}
 \label{eq:trunc_grover_filtered}
\end{equation}
where $S'$ includes filtering conditions and the EV attenuation is
\begin{equation}
  A_m = \frac{\sin^2{[(2m+1)\theta/2]}\; N - M}{N-M},
 \label{eq:Mamp_after_m}
\end{equation}
and the algorithm can be terminated whenever $A_m > A_{\textrm{th}}.$ The reduction in the number of applications of $\hat{G}$ depends on the actual threshold EV attenuation, $A_{\textrm{th}}$, and the maximum tolerable threshold EV attenuation, $A_{\text{stand}} = 1/M$, for the standard EV version~\cite{collins02}. The minimum number of applications in a truncated EV version is
\begin{equation}
 m_{\text{trunc}} \approx  m_{\text{stand}} \frac{2}{\pi} 
                         \arcsin{\sqrt{A_{\textrm{th}}/A_{\textrm{stand}} 
                        + (1-A_{\textrm{th}}/A_{\text{stand}})M/N}}
 \label{eq:final_ratio}
\end{equation}
where $m_{\text{stand}}$ is the number of applications of $\hat{G}$ required for the standard EV version~\cite{collins02}.

Whenever a quantum computing device only offers expectation value outcomes, then the truncated version of Grover's algorithm is superior to the standard version. Here the ensemble's ability to produce decisive expectation values may be used advantageously and a naive translation from the single quantum system version of an algorithm is not always optimal. The extent to which this is applicable to other algorithms is not clear and would be a worthwhile issue to investigate.

The author would like to thank Bob Griffiths for many useful discussions. This work was supported by NSF grants 9900755 and 0139974.

\end{document}